\documentclass[12pt]{article}
\usepackage{amssymb}
\def\be{\begin{equation}}
\def\ee{\end{equation}}
\def\bea{\begin{eqnarray}}
\def\eea{\end{eqnarray}}

\def\half{{\textstyle{1\over2}}}

\def\np#1{{\sl Nucl.~Phys.~\bf B#1}}
\def\pl#1{{\sl Phys.~Lett.~\bf B#1}}
\def\pr#1{{\sl Phys.~Rev.~\bf D#1}}
\def\prl#1{{\sl Phys.~Rev. Lett.~\bf #1}}

\def\cqg#1{{\sl Class.~Quant.~Grav.~\bf #1}}

\catcode`\@=11
\def\@citex[#1]#2{%
\if@filesw \immediate \write \@auxout {\string \citation {#2}}\fi
\@tempcntb\m@ne \let\@h@ld\relax \def\@citea{}%
\@cite{%
  \@for \@citeb:=#2\do {%
    \@ifundefined {b@\@citeb}%
      {\@h@ld\@citea\@tempcntb\m@ne{\bf ?}%
      \@warning {Citation `\@citeb ' on page \thepage \space undefined}}%
      {\@tempcnta\@tempcntb \advance\@tempcnta\@ne%
      \@tempcntb\number\csname b@\@citeb \endcsname \relax%
      \ifnum\@tempcnta=\@tempcntb 
        \ifx\@h@ld\relax%
          \edef \@h@ld{\@citea\csname b@\@citeb\endcsname}%
        \else%
          \edef\@h@ld{\ifmmode{-}\else--\fi\csname b@\@citeb\endcsname}%
        \fi%
      \else
        \@h@ld\@citea\csname b@\@citeb \endcsname%
        \let\@h@ld\relax%
      \fi}%
    \def\@citea{,\penalty\@highpenalty\,}%
  }\@h@ld
}{#1}}

\def\@citeb#1#2{{[#1]\if@tempswa , #2\fi}}
\def\@citeu#1#2{{$^{#1}$\if@tempswa , #2\fi }}
\def\@citep#1#2{{#1\if@tempswa , #2\fi}}

\def\bcites{         
        \catcode`\@=11
        \let\@cite=\@citeb
        \catcode`\@=12
}

\def\upcites{         
        \catcode`\@=11
        \let\@cite=\@citeu
        \catcode`\@=12
}

\def\plaincites{      
        \catcode`\@=11
        \let\@cite=\@citep
        \catcode`\@=12
}

\newcount\hour
\newcount\minute
\newtoks\amorpm
\hour=\time\divide\hour by 60
\minute=\time{\multiply\hour by 60 \global\advance\minute by-\hour}
\edef\standardtime{{\ifnum\hour<12 \global\amorpm={am}%
        \else\global\amorpm={pm}\advance\hour by-12 \fi
        \ifnum\hour=0 \hour=12 \fi
        \number\hour:\ifnum\minute<10 0\fi\number\minute\the\amorpm}}
\edef\militarytime{\number\hour:\ifnum\minute<10 0\fi\number\minute}

\def\draftlabel#1{{\@bsphack\if@filesw {\let\thepage\relax
   \xdef\@gtempa{\write\@auxout{\string
      \newlabel{#1}{{\@currentlabel}{\thepage}}}}}\@gtempa
   \if@nobreak \ifvmode\nobreak\fi\fi\fi\@esphack}
        \gdef\@eqnlabel{#1}}
\def\@eqnlabel{}
\def\@vacuum{}
\def\marginnote#1{}
\def\draftmarginnote#1{\marginpar{\raggedright\scriptsize\tt#1}}
\def\draft{
        \pagestyle{plain}
        \overfullrule=2pt
        \oddsidemargin -.75truein
        \def\@oddhead{\sl \phantom{\today\quad\militarytime} \hfil
        \smash{\Large\sl DRAFT} \hfil \today\quad\militarytime}
        \let\@evenhead\@oddhead
        \let\label=\draftlabel
        \let\marginnote=\draftmarginnote
        \def\ps@empty{\let\@mkboth\@gobbletwo
        \def\@oddfoot{\hfil \smash{\Large\sl DRAFT} \hfil}
        \let\@evenfoot\@oddhead}
        \def\@eqnnum{(\theequation)\rlap{\kern\marginparsep\tt\@eqnlabel}%
        \global\let\@eqnlabel\@vacuum}  }

\begin{document}


\hfill UTHET-03-0901

\vspace{-0.2cm}

\begin{center}
\Large
{\bf On quasi-normal modes of Kerr black holes}
\normalsize

\vspace{0.8cm}
{\bf
Suphot Musiri}\footnote{email: suphot@swu.ac.th} \\
Department of Physics, \\
Srinakharinwiroth University, Bangkok, \\
Thailand.
\vspace{.5cm}

{\bf George Siopsis}\footnote{email: gsiopsis@utk.edu}\\ Department of Physics
and Astronomy, \\
The University of Tennessee, Knoxville, \\
TN 37996 - 1200, USA.

\end{center}

\vspace{0.8cm}
\large
\centerline{\bf Abstract}
\normalsize
\vspace{.5cm}

We calculate analytically asymptotic values of quasi-normal frequencies
of four-dimensional Kerr black holes by solving the Teukolsky
wave equation.
We obtain an expression for arbitrary spin of the wave in agreement with
Hod's proposal which is based on Bohr's correspondence principle.
However, the range of frequencies is bounded from above by $1/a$, where $a$
is the angular momentum per unit mass of the black hole.
Our argument is only valid in the small-$a$ limit which includes the
Schwarzschild case.

\newpage


Quasi-normal modes of black holes in asymptotically flat space-times play an
important role in black hole physics and have attracted
a lot of attention recently~\cite{bibx1,bibx2,bibx3,bibx4,bibx5,bibx6,bibx7,bibx8,bibx9,bibx10,bibx11}.
For high overtones, the imaginary part of the quasi-normal frequencies may be
easily understood in terms of the poles of thermal Green functions,
the spacing of frequencies being $2\pi i T_H$,
where $T_H$ is the Hawking temperature.
On the other hand, the real part approaches an interesting asymptotic value
\be\label{eqi1} \Re\omega = T_H\ln 3\ee
in the case of gravitational perturbations.
This has long been known numerically~\cite{bibn1,bibn2,bibn3,bibn4,bibn5}.
The analytical expression~(\ref{eqi1}) was proposed by Hod~\cite{bibhod} and
was subsequently shown to be related to the Barbero-Immirzi parameter~\cite{bibbi1,bibbi2} of Loop Quantum Gravity (see \cite{biblqg1,biblqg2,biblqg3,biblqg4,biblqg5} and references therein).
The starting point of these arguments is
the entropy-area relation $S = \frac{1}{4G}\ A$~\cite{bibbeke},
relating the number of microstates of the black hole, $S$, to
the area of the horizon, $A$.
Mukhanov and Bekenstein~\cite{bibmb} proposed that the area spectrum of black holes be discrete with spacing of eigenvalues
\be\label{eqi2} \delta A = 4G \ln k \ \ , \ \ k = 2,3,\dots\ee
in units such that $\hbar = c = 1$.
Hod~\cite{bibhod} used Bohr's correspondence principle to relate the real part
of quasi-normal frequencies~(\ref{eqi1}) to the area spectrum~(\ref{eqi2}).
His argument suggested that $k=3$ in~(\ref{eqi2}) instead of the expected $k=2$~\cite{bibmb}.
This is intriguing
from the loop quantum gravity point of view because it suggests that the gauge group
should be $SO(3)$ rather than $SU(2)$. It seems that quasi-normal modes
may lead to a deeper understanding of black holes and quantum gravity.

The asymptotic expression~(\ref{eqi1}) has been derived analytically by Motl and Neitzke~\cite{bibx3,bibx5}
who used an interesting monodromy argument which relied heavily on the
unobservable black hole singularity.
First-order corrections have also been calculated analytically through a WKB
analysis~\cite{bibx8} for a gravitational wave and by solving the wave equation perturbatively~\cite{bibus} for arbitrary spin of the wave.

Extending the above results to rotating (Kerr) black holes does not appear to be
straightforward. By applying Bohr's correspondence principle, Hod~\cite{bibhod2} has argued that the real part of the quasi-normal frequencies of gravitational waves
ought to be given by the asymptotic expression~({\em cf.}~eq.~(\ref{eqi1}))
\begin{equation}\label{eq1k} \Re\omega = T_H \ln 3 + m\Omega \end{equation}
where $m$ is the azimuthal eigenvalue of the wave and $\Omega$ is the
angular velocity of the horizon.
On the other hand, numerical results have been obtained~\cite{bibx10} which
appear to contradict the above assertion, suggesting instead an asymptotic
expression independent of the temperature,
\begin{equation}\label{eq1kn} \Re\omega = m\Omega \end{equation}
Here, we present an analytical solution to the wave (Teukolsky~\cite{bibteu})
equation which is valid for asymptotic modes bounded from above by $1/a$,
where $a$ is the angular momentum per unit mass of the Kerr black hole.
Thus, our calculation is valid in the small-$a$ limit ($a\ll 1$), which
includes the Schwarzschild case ($a=0$). Our results confirm Hod's
expression~(\ref{eq1k}), but do not necessarily contradict the numerical result~(\ref{eq1kn}).
The latter may well be valid in the asymptotic regime $1/a \lesssim \omega$.
In the Schwarzschild limit ($a=0$), the range of frequencies considered in our
calculation extends to infinity and our expression, which generalizes~(\ref{eq1k}),
reduces to the expected form~\cite{bibx5} generalizing~(\ref{eqi1}) to
arbitrary spin of the wave.

The metric of a four-dimensional Kerr black hole may be written as
\bea ds^2 = &-& \left( 1- \frac{2Mr}{\Sigma} \right) dt^2 + \frac{4Mar\sin^2\theta}{\Sigma}\ dtd\phi + \frac{\Sigma}{\Delta}\ dr^2 + \Sigma d\theta^2
\nonumber\\
&+& \sin^2\theta \left( r^2+a^2 + \frac{2Ma^2 r\sin^2\theta}{\Sigma} \right)\
d\phi^2\eea
where
\be\label{eq6} \Sigma = r^2 + a^2 \cos^2\theta \ \ , \ \ \Delta = r^2 -2Mr + a^2 \ee
$M$ is the mass of the black hole and we have set Newton's constant $G=1$.
The roots of $\Delta$ are given by
\be\label{eqrd} r_\pm = M\pm \sqrt{M^2 - a^2} \ee
the larger being the radius of the horizon ($r_h = r_+$).
The black hole is rotating with frequency
\be \Omega = \frac{a}{2Mr_+}\ee
and its Hawking temperature is
\be\label{eqth}
T_H = \frac{1 - r_-/r_+}{8\pi M}
\ee
Small perturbations are governed by the Teukolsky wave equation~\cite{bibteu,bibteu2,bibteu3}
$$\left( \frac{(r^2+a^2)^2}{\Delta} - a^2 \sin^2\theta \right) \
\frac{\partial^2\Psi}{\partial t^2} + \frac{4Mar}{\Delta}\ \frac{\partial^2\Psi}{\partial t\partial\phi}
+ \left( \frac{a^2}{\Delta} - \frac{1}{\sin^2\theta}\right) \ \frac{\partial^2\Psi}{\partial\phi^2}
- \frac{1}{\Delta^s} \frac{\partial}{\partial r} \left( \Delta^{s+1}
\frac{\partial\Psi}{\partial r} \right)
$$
$$ - 2s \left( \frac{M(r^2-a^2)}{\Delta} - r -ia\cos\theta \right)\ \frac{\partial\Psi}{\partial t}
$$
\be\label{eq5a} - \frac{1}{\sin\theta} \frac{\partial}{\partial\theta}
\left( \sin\theta \frac{\partial\Psi}{\partial\theta} \right)
-2s \left( \frac{a(r-M)}{\Delta} + \frac{i\cos\theta}{\sin^2\theta} \right)\
\frac{\partial\Psi}{\partial\phi} + (s^2\cot^2\theta - s)\Psi = 0\ee
where $s= 0, -1, -2$ for scalar, electromagnetic and gravitational wave, respectively.

Separating variables
\be \Psi = e^{-i\omega t} e^{im\phi} S(\theta) f(r) \ee
we obtain an angular and a radial equation, respectively,
\be\label{eqang} \frac{1}{\sin\theta} (\sin\theta\ S')' +
\left( a^2\omega^2 \cos^2\theta - \frac{m^2}{\sin^2\theta} -2a\omega s\cos\theta
- \frac{2ms\cos\theta}{\sin^2\theta} - s^2 \cot^2\theta \right) S = -(A+s)S\ee
\be\label{eq5ar} \frac{1}{\Delta^s} (\Delta^{s+1} f')' + V(r) f = (A+a^2\omega^2) f\ee
where the potential is given by
\be\label{eqV} V(r) = \frac{(r^2+a^2)^2\omega^2 - 4aMr\omega m + a^2m^2 +2ia(r-M) ms -2iM (r^2 - a^2) \omega s}{\Delta} + 2ir\omega s \ee
and $A$ is an eigenvalue to be determined by solving the angular eq.~(\ref{eqang}).
Before attempting to solve these equations, let us simplify them by placing the horizon
at $r=1$. This is accomplished by setting
\be 2M = 1+a^2 \ee
Then the roots of $\Delta$ (eq.~(\ref{eqrd}))
are given by
\be r_- = a^2 \ \ , \ \ r_+ = 1 \ee
respectively.
Let us try to solve the two wave equations~(\ref{eqang}) and (\ref{eq5ar}) by expanding in $a$.
We shall keep terms up to $o(a)$. Moreover, we shall assume that
the frequency $\omega$ is large but bounded from above by the large
parameter $1/a$,
\be \omega \lesssim 1/a \ee
Thus, $\omega$ is not in the asymptotic regime but rather in an intermediate
range. This range of frequencies includes the asymptotic regime in the Schwarzschild limit $a\to 0$.
Our calculation is valid in the limit of small $a$ ($a\ll 1$).

The solutions of the angular equation~(\ref{eqang}) to lowest order are spin-weighted spherical
harmonics~\cite{bibteu} and the separation constant (eigenvalue) $A$ is
\be A = \ell(\ell+1) - s(s+1) + o(a\omega)\ee
It is convenient to express the radial eq.~(\ref{eq5ar}) in terms of a ``tortoise coordinate.''
To define it, observe that near the horizon ($r\to 1$), the wavefunction behaves as
\be\label{eq19} f(r) \sim (r-1)^\lambda \ \ , \ \ \lambda = i(\omega - am) + o(1/\omega)\ee
On the other hand, at infinity ($r\to\infty$), we have
\be f(r) \sim e^{i\omega r} \ee
By introducing the variable (``tortoise coordinate'')
\be\label{eqtor} z = \omega r + (\omega - am) \ln (r-1) \ee
we may express the boundary conditions simply as
\be f(z) \sim e^{\pm iz} \ \ , \ \ z\to \pm\infty
\ee
Moreover, the monodromy for the function $e^{iz} f(z)$ (which approaches
a constant as $r\to\infty$ (i.e., $z\to +\infty$)) around the singular point
$r=1$ to this order ($o(a)$) is easily deduced from~(\ref{eq19}),
\be\label{eqmono} \mathcal{M} (1) = e^{4\pi (\omega - am)} \ee
This will serve as the definition of the boundary condition at the horizon~\cite{bibx5}.
The contour surrounding the singularity $r=1$ can be deformed in the
complex $r$-plane so that it either lies beyond the horizon ($\Re r
< 1$) or at infinity ($r\to \infty$).
Then the monodromy only gets a contribution from the segment lying beyond the
horizon.
To express the radial equation in terms of the tortoise coordinate, define
\be f(r) = \Delta_0^{-s/2}\ \frac{R(r)}{\sqrt{r(\omega r-am)}} \ee
where $\Delta_0 = r(r-1)$ (note from eq.~(\ref{eq6}) that $\Delta = \Delta_0 +
o(a^2)$).
Inverting~(\ref{eqtor}),
\be\label{eqtora} r = \sqrt{-\frac{2z}{\omega}} + o(1/\omega) \ee
we obtain the radial equation~(\ref{eq5ar})
to lowest order in $1/\sqrt\omega$ in terms of $R$,
\be\label{eq5b} \frac{d^2R}{dz^2} + \left\{ 1 + \frac{3is}{2z} + \frac{4-s^2
- 4iams}{16z^2} \right\}\ R = 0\ee
to be solved along the entire real axis.
This is Whittaker's equation~\cite{bibgr}.
Two linearly independent solutions are (setting $x=2iz$)
\be M_{\kappa,\pm\mu}(x) = e^{-x/2} x^{\pm\mu + 1/2} M(\half \pm \mu - \kappa, 1\pm 2\mu, x)\ee
where
\be\label{eqxi} \kappa = \frac{3s}{4}\ \ , \ \ \mu^2 = \frac{s(s+4iam)}{16}\ee
and $M$ is Kummer's function (also called $\Phi$).
For our purposes, the following linear combination (Whittaker's function) will be useful:
\be
W_{\kappa,\mu}(x) =
\frac{\Gamma(-2\mu)}{\Gamma(\half - \mu-\kappa)}\ M_{\kappa,\mu}(x)
+ \frac{\Gamma(2\mu)}{\Gamma(\half +\mu-\kappa)}\ M_{\kappa,-\mu}(x)
\ee
due to its clean asymptotic behavior,
\be W_{\kappa,\mu}(x) \sim e^{-x/2}\ x^\kappa\ (1 + o(1/x)) \ee
as $|x|\to\infty$.

To compute the monodromy around the singularity $r=1$, we shall deform the
contour so that it gets mapped onto the real axis in the $z$-plane. Near the
singularity $z=0$, we have $z \approx -\frac{\omega}{2r_0}\, r^2$ (eq.~(\ref{eqtora})).
We shall choose our contour in the complex $r$ plane so that near $r=0$,
the positive real axis and the negative real axis in the $z$-plane are
mapped onto
\be \arg r = \pi - \frac{\arg\omega}{2} \ \ , \ \ \arg r = \frac{3\pi}{2}
- \frac{\arg\omega}{2} \ee
in the $r$-plane.
These two segments form a $\pi /2$ angle (independent of $\arg\omega$). To avoid the $r=0$ singularity,
we shall go around an arc of angle $3\pi /2$. This translates to an angle
$3\pi$ around $z=0$ in the $z$-plane~\cite{bibx5}.
To implement this analytic continuation,
observe
\bea M_{\kappa,\pm\mu}(e^{3\pi i} x) &=& e^{3\pi i (\pm\mu + 1/2)}\
e^{-x/2} x^{\pm\mu + 1/2} M(\half \pm \mu + \kappa, 1\pm 2\mu, x)\nonumber \\
&=& -i e^{\pm 3\pi i\mu} M_{-\kappa, \pm\mu} (x)\eea
where we used
\be M(a,b,-x) = e^{-x} M(b-a,b,x) \ee
Therefore,
\be
W_{\kappa,\mu}(e^{3\pi i} x) =
-i e^{3\pi i \mu}\ \frac{\Gamma(-2\mu)}{\Gamma(\half - \mu-\kappa)}\ M_{-\kappa,\mu}(x)
-i e^{-3\pi i \mu}\ \frac{\Gamma(2\mu)}{\Gamma(\half +\mu-\kappa)}\ M_{-\kappa,-\mu}(x)
\ee
For the asymptotic behavior, we need
\be M_{-\kappa,\mu} (x) = \frac{\Gamma(1+2\mu)}{\Gamma(\half +\mu +\kappa)}\
e^{-i\pi\kappa} W_{\kappa,\mu} (e^{i\pi} x) +
\frac{\Gamma(1+2\mu)}{\Gamma(\half +\mu -\kappa)}\
e^{-i\pi(\half+\mu +\kappa)} W_{-\kappa,\mu} (x)
\ee
As $|x|\to\infty$, we obtain
\be M_{-\kappa,\mu} (x) \sim \frac{\Gamma(1+2\mu)}{\Gamma(\half +\mu +\kappa)}\
e^{-i\pi\kappa}\ e^{x/2}\ (-x)^\kappa +
\frac{\Gamma(1+2\mu)}{\Gamma(\half +\mu -\kappa)}\
e^{-i\pi(\half+\mu +\kappa)} e^{-x/2}\ x^{-\kappa}
\ee
and so
\be W_{\kappa,\mu}(e^{3\pi i} x) \sim \mathcal{A} e^{x/2} x^\kappa
+ \mathcal{B} e^{-x/2} x^{-\kappa} \ee
where
\be\label{eqA} \mathcal{A} =
-i e^{3\pi i \mu}\ \frac{\Gamma(-2\mu)}{\Gamma(\half - \mu-\kappa)}\
\frac{\Gamma(1+2\mu)}{\Gamma(\half +\mu +\kappa)}\ e^{-\pi i \kappa}
+ (\mu\to -\mu)
\ee
\be \mathcal{B} =
-i e^{3\pi i \mu}\ \frac{\Gamma(-2\mu)}{\Gamma(\half - \mu-\kappa)}\
\frac{\Gamma(1+2\mu)}{\Gamma(\half +\mu -\kappa)}\
e^{-i\pi(\half+\mu +\kappa)} + (\mu\to -\mu)
\ee
Using~(\ref{eqxi}), after some algebra we obtain from~(\ref{eqA})
\be\label{eqf1} \mathcal{A} = - (1+2\cos \pi s) + o(a^2)\ee
where we also used the identities
\be \Gamma (1-x)\Gamma(x) = \frac{\pi}{\sin\pi x} \ \ , \ \
\Gamma(\half+x)\Gamma(\half -x) = \frac{\pi}{\cos\pi x} \ee
Eq.~(\ref{eqf1}) gives the correct Schwarzschild limit~\cite{bibx5}.
Notice that there are no $o(a)$ corrections.
Using~(\ref{eqmono}), we obtain for the monodromy around $r=1$,
\be \mathcal{M} (1) = e^{4\pi(\omega -ma)} = \mathcal{A}
\ee
therefore,
\be\label{eq40} \Re\omega = \frac{1}{4\pi}\ \ln (1+2\cos \pi s) + ma + o(a^2) \ee
which is in agreement with Hod's formula~(\ref{eq1k}) in the case of gravitational waves ($s=-2$) and in the
small-$a$ limit (in which $\Omega \approx a$; also note $T_H \approx \frac{1}{4\pi}$ in our units).
However, it should be emphasized that these are not asymptotic values of
quasi-normal modes being bounded from above by $1/a$.

To summarize, we have obtained an analytical expression for asymptotic values
of quasi-normal frequencies of Kerr black holes by solving the wave (Teukolsky~\cite{bibteu})
equation perturbatively.
The zeroth-order approximation is Whittaker's equation. We applied the
monodromy argument of Motl and Neitzke~\cite{bibx5} to the zeroth-order
solution and arrived at an explicit expression for arbitrary spin of the wave~(eq.~(\ref{eq40})).
This result is in agreement with Hod's suggestion~(\ref{eq1k}) based on Bohr's
correspondence principle in the case of gravitational waves and in the
small-$a$ limit.
Eq.~(\ref{eq40}) is only applicable to modes bounded from above by $1/a$.
It would be interesting to extend our results to general values of $a$
and an unbounded asymptotic regime of quasi-normal modes, $1/a \lesssim \omega$.
Such an extension is far from straightforward.

\section*{Acknowledgments}

We wish to thank J.~D.~Bekenstein, E.~Berti, V.~Cardoso, K.~Kokkotas, L.~Motl and A.~Neitzke
 for useful discussions.
G.~S.~is supported by the US Department of Energy under grant
DE-FG05-91ER40627.

\end{document}